\documentclass[preprint,noshowkeys,amsmath,amssymb,superscriptaddress,nofootinbib]{revtex4}

\usepackage{bm}
\usepackage{slashed}

\usepackage[pdftex]{graphicx,color}
\usepackage{epstopdf}
\usepackage[bookmarks=true,bookmarksnumbered=true,bookmarkstype=toc, colorlinks=true, 
citecolor=blue, linkcolor=blue]{hyperref} 
\usepackage{units}

\newcommand{\lsim}{\raise0.3ex\hbox{$\;<$\kern-0.75em\raise-1.1ex\hbox{$\sim\;$}}}
\newcommand{\gsim}{\raise0.3ex\hbox{$\;>$\kern-0.75em\raise-1.1ex\hbox{$\sim\;$}}}
\usepackage{mathtools}% http://ctan.org/pkg/mathtools
\usepackage{color}

\definecolor{green}{cmyk}{1,0,1,0}

\definecolor{pink}{cmyk}{0,0.5,0,0}

\definecolor{pastelpink}{cmyk}{0,0.25,0,0}

\definecolor{softpink}{cmyk}{0,0.125,0,0}

\definecolor{purple}{cmyk}{0.5,1.0,0.1,0}

\definecolor{violet}{cmyk}{0.75,1,0.25,0}

\usepackage{ulem}  % \sout{ text }
\newcommand{\Beex}{${^8}$Be$^\ast$ }    % excited state of 8Be

\newcommand{\Begr}{${^8}$Be }                % grond state of 8Be

\newcommand{\mBeex}{ {^8} \mathrm{Be}^\ast}
\newcommand{\mBegr}{ {^8} \mathrm{Be}}

\usepackage[utf8]{inputenc}

\preprint{UME-PP-012}
\preprint{EPHOU-20-007}

\begin{document}

%%%%% declaration for front matter%%%%%%%%%%%%%%%%%%%%%%%%%%%%%%%%%%%
\title{Atomki anomaly in gauged $U(1)_R$ symmetric model}

\author{Osamu Seto}
\email{seto@particle.sci.hokudai.ac.jp}
\affiliation{Institute for the Advancement of Higher Education, Hokkaido University, Sapporo 060-0817, Japan}
\affiliation{Department of Physics, Hokkaido University, Sapporo 060-0810, Japan}

\author{Takashi Shimomura}
\email{shimomura@cc.miyazaki-u.ac.jp}
\affiliation{Faculty of Education, Miyazaki University, Miyazaki, 889-2192, Japan}

%%%%% typeset front matter (including abstract) %%%%%%%%%%%%%%%%%%%%%%
\begin{abstract}
The  Atomki collaboration has reported that unexpected excesses have been observed in the rare decays 
of Beryllium nucleus. It is claimed that such excesses can suggest the existence of a new boson, called $X$, 
with the mass of about $17$ MeV. 
To solve the Atomki anomaly, we consider a model with gauged $U(1)_R$ symmetry and identify  
the new gauge boson with the $X$ boson. 
We also introduce two $SU(2)$ doublet Higgs bosons and one singlet Higgs boson, and discuss 
a very stringent constraint from neutrino-electron scattering. 
It is found that the $U(1)_R$ charges of the doublet scalars are determined to evade the constraint.
In the end, we find the parameter region in which the Atomki signal and all experimental constraints 
can be simultaneously satisfied.
\end{abstract}

\date{\today}

\maketitle

%%%%%%%%%%%%%%%%%%%
\section{Introduction}
%%%%%%%%%%%%%%%%%%%
The Atomki collaboration has been reporting results that unexpected excesses were found in the Internal Pair Creation 
(IPC) decay of Beryllium (Be) \cite{Krasznahorkay:2015iga, Krasznahorkay:2017qfd,Krasznahorkay:2017gwn,
Krasznahorkay:2017bwh,Krasznahorkay:2018snd} and Helium (He) \cite{Krasznahorkay:2019lyl,Firak:2020eil} nuclei. 
In the reports, the excesses appear as bumps in the distributions 
of the invariant mass and opening angle of an emitted positron ($e^+$) and electron ($e^-$) pair from the IPC decays of \Beex and  $^4$He, 
\begin{align}
\mBeex(18.15~\mathrm{MeV}) \to \mBegr + e^+ + e^- \label{eq:Be-decay}, \\
^4 \mathrm{He}(21.01~\mathrm{MeV}) \to  ^4 \mathrm{He }+ e^+ + e^-,
\end{align}
respectively.
These bumps seem not to be explained within the standard nuclear physics \cite{Zhang:2017zap}, 
even if parity violating decays are taken into account. 
The collaboration reported that the bumps can be well fitted simultaneously under the assumption that 
a hypothetical boson $X$ with the mass of $17.01 \pm 0.16$ and $19.68 \pm 0.25$ MeV is 
produced through \Beex 
and $^4$He decays, followed by $X$ decay into a $e^+$-$e^-$ pair, respectively.
Such a light boson does not exist in the Standard Model (SM) of particle physics. Therefore, the anomaly can be 
considered as a signal of new physics beyond the SM.

The hypothetical boson $X$, in principle, can be a vector, axial-vector, scalar and pseudo-scalar boson. 
Among these possibilities, the scalar boson hypothesis is discarded due to the conservation of angular momentum 
in the decay  Eq.~\eqref{eq:Be-decay}
\cite{Feng:2016jff,Feng:2016ysn}.
Vector boson hypothesis was firstly studied in \cite{Feng:2016jff,Feng:2016ysn} in a 
gauged $B-L$ symmetric model, taking various experimental constraints into account. 
Then, many models have been proposed in contexts of 
an extra U(1) gauge symmetry \cite{Gu:2016ege,Neves:2017rcn,Pulice:2019xel}, 
dark matter \cite{Kitahara:2016zyb,Jia:2017iyc,Jia:2018mkc}, 
neutrino physics \cite{Seto:2016pks}, 
lepton anomalous magnetic moments \cite{BORDES:2019wcp,Kirpichnikov:2020tcf, Hati:2020fzp} 
and others \cite{Neves:2018bay,Nam:2019osu,Wong:2020hjc,Tursunov:2020wfy,Chen:2020arr}.
Experimental searches of the $X$ boson are also studied in \cite{Alikhanov:2017cpy,Jentschura:2018zjv,Koch:2020ouk,Jentschura:2020zlr}.
In \cite{Feng:2016ysn}, it was shown that there are two restrictive constraints to explain the Atomki anomaly. 
The first constraint comes from the a rare decay of neutral pion, $\pi^0 \to \gamma X$, measured by the NA48/2 experiment. 
This constraint sets a very stringent bound on the coupling of $X$ to proton because the decay branching ratio of the rare decay 
is scaled by the proton coupling. From this fact, such a vector boson is named as a protophobic boson. 
The second constraint comes from neutrino-electron scattering measured by the TEXONO experiment. 
It is difficult to evade this constraint and neutral pion constraint simultaneously. 
Therefore, new leptonic states are introduced to evade this constraint in \cite{Feng:2016ysn}, 
or no interaction of the $X$ boson to active neutrinos is ad hoc assumed. 

An axial-vector boson hypothesis also has been studied in \cite{Kozaczuk:2016nma}. 
This hypothesis has two advantages. One is that the constraints from neutral pion decay can be easily evaded 
because the decay receives no contribution from the axial anomaly. The other advantage is that the partial decay width 
of \Beex is proportional to $k_X$,\footnote{There is also $k_X^3$ term in the partial width. Following the discussion in 
 \cite{Kozaczuk:2016nma}, we neglected that term, which would be suppressed  because $k_X$ is smaller than mass of 
 the $X$ boson.} the $X$'s three momentum, while it is proportional to $k_X^3$ in a vector boson hypothesis. 
Because of this momentum dependence, coupling constants of $X$ to quarks can be much weaker to explain 
the Atomki anomaly than that in the vector-boson hypothesis case. 
Then, it is possible to evade several experimental constraints in the axial-vector boson hypothesis.
Several models with axial-vector boson have been proposed in \cite{DelleRose:2017xil,DelleRose:2018eic,DelleRose:2018pgm}. 
In spite of these advantages, the constraint from neutrino-electron scattering is still very stringent and 
requires to suppress neutrino couplings to $X$. In \cite{Kozaczuk:2016nma}, it is assumed that neutrino 
couplings to the $X$ boson vanish, and in \cite{DelleRose:2017xil,DelleRose:2018eic,DelleRose:2018pgm}, many fermions are introduced to cancel the neutrino 
couplings.  
In the end, pseudo-scalar hypothesis was studied in \cite{Ellwanger:2016wfe}.
Decay widths of these three hypotheses are found in \cite{Feng:2020mbt}.

In this work, we pursue the axial-vector hypothesis and consider a $U(1)_R$ gauge symmetry \cite{Jung:2009jz} 
where the gauge boson is identified with the $X$ boson.  The $U(1)_R$ gauge symmetry is defined that  
only right-handed fermions are charged while left-handed ones are not charged. 
Then, the $U(1)_R$ gauge boson has both vectorial and axial interactions to fermions. 
The existence of the axial interaction allows coupling constants to be weaker to satisfy the Atomki signal. 
With weaker couplings, a contribution to neutral pion decay from vectorial interactions is much suppressed. 
It was shown in \cite{Ko:2012hd, Ko:2013zsa} that flavour changing neutral currents can be suppressed due to 
$U(1)_R$ symmetry in two Higgs doublet extension. It was also shown in 
\cite{Nomura:2017ezy,Nomura:2017tih,Nomura:2018mwr,Chao:2017rwv} that neutrino masses and mixing, dark matter and 
the muon anomalous magnetic moment can be explained in models with the $U(1)_R$ gauge symmetry.
Motivated by these previous works, we construct a minimal model to explain the Atomki anomaly with $U(1)_R$ gauge symmetry.

This paper is organized as follows. In Sec.~\ref{sec:model}, we introduce our model as a minimal setup to explain the Atomki anomaly.
In Sec.~\ref{sec:coupling}, we give the coupling constants of fermions to the $X$ boson and show the allowed region of gauge coupling 
constant. Then, the signal requirement and experimental constraints are explained in Sec.~\ref{sec:signal-constraint} and our numerical 
results are shown in Sec.~\ref{sec:num-results}. In the end, we give our conclusion in Sec.~\ref{sec:conclusion}.

%%%%%%%%%%%%%%%%%%%
%%%%%%%%%%%%%%%%%%%
\section{Model} \label{sec:model}
%%%%%%%%%%%%%%%%%%%
%%%%%%%%%%%%%%%%%%%
We start our discussion by introducing our model. The gauge symmetry of the model is defined as 
$G_{\mathrm{SM}} \times U(1)_R$, where $G_{\mathrm{SM}}$ stands for the gauge symmetry of the SM.
Under the $U(1)_R$ gauge symmetry, right-handed chiral fermions are charged while left-handed chiral ones  
are singlet \cite{Nomura:2017tih}. 
Only with the SM matter content, such a charge assignment generally leads to non-vanishing gauge-anomalies due to 
$U(1)_R$ current contributions.
 Therefore, new fermions charged under $U(1)_R$ must be introduced to cancel the gauge anomalies. 
One of the simplest solutions for non-vanishing anomalies is to add three right-handed fermions,  $N_i~(i=1,2,3)$, 
which are singlet under the SM gauge symmetries. 
The charge assignment of the fermions in our model is shown in Table \ref{tab:matter-contents}. 
In the Table, $SU(3),~SU(2)_L$ and $U(1)_Y$ represent the SM strong, weak and hypercharge gauge groups. 
The symbols, $Q$ and $u_R$, $d_R$ represent left-handed quarks and right-handed up-type, down-type quarks, respectively, 
and $L$ and $e_R$ represent left-handed leptons and right-handed charged leptons, respectively. 
Without loss of generality, we can fix the $U(1)_R$ charge of $u_R$ to $+\frac{1}{2}$ as the overall normalization. 
Then, the gauge charges of the other fermions are determined from anomaly-free conditions as shown in the Table \ref{tab:matter-contents}.
For the Higgs field $H_1$, we assign its $U(1)_R$ charge to $q_1$, which can not be determined from anomaly-free conditions. 
However, if requiring for the model to be minimal, $q_1$ should be taken as $+\frac{1}{2}$ so that quarks 
and charged leptons can form Yukawa interactions with $H_1$ in the same manner of the SM. 
Furthermore, with this charge assignment, left-handed neutrinos can form the Yukawa interaction with $N_i$. 
Therefore, we identify $N_i$ as right-handed neutrinos in the following discussions.

To explain the Atomki anomaly, we further extend the matter content by adding a $SU(2)_L$ doublet scalar field $H_2$ 
and a $SU(2)_L$ singlet scalar field $S$. 
Firstly, it is shown in \cite{Heeck:2014zfa} that neutrinos can not be Dirac particle due to the constraints 
from $\Delta N_{\mathrm{eff}}$ unless the coupling constant of neutrinos are extremely small. 
This constraint can be avoided when right-handed neutrinos have  Majorana masses. 
The $SU(2)_L$ singlet scalar field is introduced to give a mass to the $X$ boson and  Majorana masses to $N_i$ after 
spontaneous breaking of $U(1)_R$. Thus its $U(1)_R$ charge is assigned to $-1$.
The new $SU(2)_L$ doublet scalar field is also introduced. It plays an important role to reduce the mixing between 
left-handed neutrinos and the $U(1)_R$ gauge boson, $X$, so that the stringent constraint from neutrino-electron scattering 
is avoided. 
The $U(1)_R$ charge of $H_2$ is arbitrary, and we will discuss possible charge assignments later. The charge 
assignment of the new scalars is also shown in Table \ref{tab:matter-contents}, where we denote the $U(1)_R$ charge 
of $H_2$ as $q_2$.

%%%%%%%%%%%%%%%%%%%
\subsection{Lagrangian}
%%%%%%%%%%%%%%%%%%%

%%%%%%%%%%%%%%%%%%%%%%%%%%%%%%%%%%%%%%%
\begin{table}[t]
  \begin{center}
    \begin{tabular}{|c||cccccc||ccc|} \hline
	& ~~~$Q$~~~ & ~~~$u_R$~~~ & ~~~$d_R$~~~ & ~~~$L$~~~ & ~~~$e_R$~~~ & 
	~~~$N$~~~ & ~~~$H_1$~~~ & ~~~$H_2$~~~ & ~~~$S$~~~  \\ \hline \hline
	$SU(3)$ & $3$ & $3$ & $3$ & $1$ & $1$ & $1$ & $1$ & $1$ & $1$ \\ \hline
	$SU(2)_L$ & $2$ & $1$ & $1$ & $2$ & $1$ & $1$ & $2$ & $2$  & $1$ \\ \hline
	$U(1)_{Y}$ & $\frac{1}{6}$ & $+\frac{2}{3}$ & $-\frac{1}{3}$ & $-\frac{1}{2}$ & $-1$ & $0$ 
	     & $+\frac{1}{2}$ & $+\frac{1}{2}$ & $0$\\ \hline
	$U(1)_R$ & $0$ & $+\frac{1}{2}$ & $-\frac{1}{2}$ & $0$ & $-\frac{1}{2}$ & $+\frac{1}{2}$ & $q_1 = +\frac{1}{2}$ & $q_2$ & $-1$ \\ \hline
    \end{tabular}
  \end{center}
\caption{Matter contents and charge assignment of the model. }
\label{tab:matter-contents}
\end{table}
%%%%%%%%%%%%%%%%%%%%%%%%%%%%%%%%%%%%%%%

The Lagrangian of the model takes the form of 
\begin{align}
\mathcal{L} &= \mathcal{L}_{\mathrm{fermion}} + \mathcal{L}_{\mathrm{scalar}} + \mathcal{L}_{\mathrm{gauge}} + \mathcal{L}_{\mathrm{yukawa}} - V,
\end{align}
where each term denotes the fermion, scalar, gauge and Yukawa sector Lagrangian which are defined as 
\begin{subequations}
\begin{align}
\mathcal{L}_{\mathrm{fermion}} &= i \sum_f \overline{f} \slashed{D} f, \label{eq:lag-fermion}\\ 
\mathcal{L}_{\mathrm{scalar}} &= |D_\mu H_1|^2 + | D_\mu H_2|^2 + |D_\mu S|^2, \label{eq:lag-scalar} \\
\mathcal{L}_{\mathrm{gauge}} &= -\frac{1}{4} \tilde{W}_{\mu \nu} \tilde{W}^{\mu \nu} - \frac{1}{4} \tilde{B}_{\mu \nu} \tilde{B}^{\mu \nu} 
-\frac{1}{4} \tilde{X}_{\mu \nu} \tilde{X}^{\mu \nu}  + \frac{\epsilon}{2} \tilde{B}_{\mu \nu} \tilde{X}^{\mu \nu}, \label{eq:lag-gauge}\\
\mathcal{L}_{\mathrm{yukawa}} &= Y_u \overline{Q} \tilde{H_1} u_R + Y_d \overline{Q} H_1 d_R  + Y_e \overline{L} H_1 e_R \nonumber \\
&\quad + Y_\nu \overline{L} \tilde{H_1} N  + Y_N \overline{N^c} S N + h.c. \label{eq:lag-yukawa}
\end{align}
\label{eq:lagrangian}
\end{subequations}
and $V$ is the scalar potential which is given below.
In Eqs.~\eqref{eq:lagrangian}, $f$ represents the fermions ($Q,~u_L,~u_R$ and $L,~e_R,~N$), 
and $\tilde{W},~\tilde{B}$ and $\tilde{X}$ represent 
the gauge fields and their field strengths in the interaction basis of $SU(2)_L,~U(1)_Y$ and $U(1)_R$, respectively. 
The covariant derivative in Eqs.~\eqref{eq:lag-fermion} and \eqref{eq:lag-scalar} is given by 
\begin{align}
D_\mu = \partial_\mu - i g_2 \tilde{W}_\mu - i Y  g_1 \tilde{B} - i x g' \tilde{X}_\mu,
\end{align}
where $Y$ and $x$ represent the $U(1)_Y$ and $U(1)_R$ charges of each particle. The gauge coupling constants of 
$SU(2)_L.~U(1)_Y$ and $U(1)_R$ are denoted as $g_2,~g_1$ and $g'$, respectively.
In Eq.~\eqref{eq:lag-gauge}, the gauge symmetry of the model allows the gauge kinetic mixing term 
between $\tilde{B}$ and $\tilde{X}$, and its magnitude is parameterized by the constant parameter $\epsilon$.
In Eq.~\eqref{eq:lag-yukawa}, the Dirac Yukawa matrices are denoted as $Y_u,~Y_d$ and $Y_e,~Y_\nu$ 
for up, down quarks and charged leptons, neutrinos, respectively. 
The Yukawa matrix for right-handed neutrinos is denoted as $Y_N$. 
Here $\tilde{H_1}$ represents $i\sigma_2 H_1^\ast$ where $\sigma_2$ is the Pauli matrix. 
Note that flavour and generation indices are omitted for simplicity.

The scalar potential $V$ can be divided into two parts. One consists of the terms independent of the $U(1)_R$ charge 
assignment of $H_2$, and the other consists of the terms dependent on that.
The charge-independent part, $V_0$, is given by
\begin{align}
V_0 &= -\mu_1^2 |H_1|^2 -\mu_2^2 |H_2|^2 -\mu_s^2 |S|^2 
+ \frac{\lambda_1}{2} |H_1|^4 + \frac{\lambda_2}{2} |H_2|^4 + \frac{\lambda_s}{2} |S|^4 \nonumber \\
&\quad + \lambda'_1 |H_1^\dagger H_2|^2 + \lambda'_2 |H_1|^2 |H_2|^2 
+ \lambda'_3 |S|^2 |H_1|^2 + \lambda'_4 |S|^2 |H_2|^2,
\end{align}
where we assume the mass parameters as well as the quartic couplings to be positive so that spontaneous breaking 
of the symmetries successfully occurs, and no runaway directions appear in the potential. 
With the above potential, we obtain five Nambu-Goldstone bosons after $H_1,~H_2$ and $S$ develop 
vacuum expectation values (VEVs). Two of those are absorbed by the charged weak boson, $W^\pm$, and other two are absorbed 
by the neutral weak boson $Z$ and the new gauge boson, $\tilde{X}$. Then, one massless CP-odd scalar remains in the spectrum, 
which corresponds to the broken degree of freedom of the phase rotation of $H_2$. 
Such a massless scalar boson causes serious problems by carrying the energy of stars and conflicts with meson decay 
measurement such as an axion does \cite{Kim:2008hd, Kawasaki:2013ae}.
Therefore we need to introduce other interaction terms which give the mass to the CP-odd scalars after the symmetry breaking. 
In this sense, a possible choice of $q_2$ is determined. We classify models with different choices of $q_2$ given in 
Table \ref{tab:model}.
%%%%%%%%%%%%%%%%%%%%%%%%%%%%%%%%%%%%%%%
\begin{table}[t]
  \begin{center}
    \begin{tabular}{|c|c|} \hline
                  & ~~~~~~~$q_2$~~~~~~~ \\ \hline 
    ~~~Model 1~~~ &  ~~$-1/2$~~ \\ \hline
    ~~~Model 2~~~ &  ~~$+3/2$~~ \\ \hline
    ~~~Model 3~~~ &  ~~$-3/2$~~ \\ \hline
    ~~~Model 4~~~ &  ~~$+5/2$~~ \\ \hline
    \end{tabular}
  \end{center}
\caption{The charge assignments of $H_2$ for each model.}
\label{tab:model}
\end{table}
%%%%%%%%%%%%%%%%%%%%%%%%%%%%%%%%%%%%%%%

The charge-dependent scalar potential in each model is given by
\begin{subequations}
\begin{align}
\mathrm{Model~1}:&~~\Delta V_1 = A_1 S H_2^\dagger H_1 + h.c., \\
\mathrm{Model~2}:&~~\Delta V_2 = A_2 S H_1^\dagger H_2 + h.c., \\
\mathrm{Model~3}:&~~\Delta V_3 = \kappa_1 S^2 H_2^\dagger H_1 + h.c., \\
\mathrm{Model~4}:&~~\Delta V_4 = \kappa_2 S^2 H_1^\dagger H_2 + h.c., 
\end{align}
\label{eq:delta-V}
\end{subequations}
where the parameters, $A_{1,2}$ and $\kappa_{3,4}$, can be taken real by using phase rotation of $H_2$. 
One example of the parameter sets to reproduce the Higgs mass, $125$ GeV, for Model $1$ is found as
\begin{align}
\begin{split}
v &= 246.0~\mathrm{GeV},~~v_s = v,~~\cos2\beta = 0.1~~(\tan\beta = 0.9045),  \\
\lambda_1 &= 0.7,~~\lambda_2 = \lambda_s = 1.0,~~\lambda'_1 = \lambda'_2 = 0,  \\
\lambda'_3 &= \lambda'_4 = 5.0,  \\
A_1 &= -\sqrt{2} \lambda'_3 \tan\beta v_s + 80~\mathrm{GeV}. 
\end{split}
\label{eq:param-example}
\end{align}
With these parameters, the Higgs couplings to the weak gauge bosons are the same as those of the SM, 
and the coupling to the $X$ boson vanishes. The messes of other extra Higgs scalars are also large enough.
However, details of the scalar sector is essentially irrelevant 
for our study about the Atomki anomaly. Therefore, in the following discussions, 
we assume that the parameters in the scalar potential are appropriately chosen so that the new gauge boson acquires 
the mass required to explain the Atomki anomaly.

%%%%%%%%%%%%%%%%%%%%%%%%%%%%%%%%%%%%%%%%%%%%%%
\subsection{Gauge boson Masses and Mass Eigenstates} \label{subsec:gauge-sector}
%%%%%%%%%%%%%%%%%%%%%%%%%%%%%%%%%%%%%%%%%%%%%%
After the EW and $U(1)_R$ symmetries are broken down, the gauge boson masses are generated via 
the VEVs of the scalar fields.  We denote the VEVs as
\begin{align}
\langle H_1 \rangle = \frac{1}{\sqrt{2}}
	\begin{pmatrix}
	0 \\ v_1
	\end{pmatrix},~~
\langle H_2 \rangle = \frac{1}{\sqrt{2}}
	\begin{pmatrix}
	0 \\ v_2
	\end{pmatrix},~~
\langle S \rangle = \frac{1}{\sqrt{2}} v_s,
\label{eq:vevs}
\end{align}
and each scalar field is expanded around its VEV as 
\begin{align}
H_1 = 
	\begin{pmatrix}
	H_1^+ \\ 
	\frac{1}{\sqrt{2}} (v_1 + h_1 + i a_1)
	\end{pmatrix},~~
H_2 = 
	\begin{pmatrix}
	H_2^+ \\ 
	\frac{1}{\sqrt{2}} (v_2 + h_2 + i a_2)
	\end{pmatrix},~~
S = \frac{1}{\sqrt{2}} (v_s + s + i \zeta).
\label{eq:scalar-fields}
\end{align}

Then, the mass terms of the gauge fields are given by
\begin{align}
\mathcal{L}_{\mathrm{gauge, mass}} &= 
\frac{1}{8} \sum_{i=1}^2 v_i^2 \left[
	2 g_2^2 W^+_\mu W^{-\mu} 
	 + ( - \sqrt{g_1^2 + g_2^2} \tilde{Z}_\mu + 2 q_i g' \tilde{X}_\mu)^2
	\right]  
+ \frac{1}{2} g'^2 v_s^2 \tilde{X}_\mu \tilde{X}^\mu,
\end{align}
with 
\begin{subequations}
\begin{align}
W^\pm_\mu &= \frac{1}{\sqrt{2}} (\tilde{W}^1_\mu \mp i \tilde{W}^2_\mu), \\
\tilde{Z}_\mu &= \cos\theta_W \tilde{W}^3_\mu - \sin\theta_W \tilde{B}_\mu,\\
\tilde{A}_\mu &= \sin\theta_W \tilde{W}^3_\mu + \cos\theta_W \tilde{B}_\mu.
\end{align}
\label{eq:sm-gauge-bosons}
\end{subequations}
Here, $\theta_W$ is the Weinberg angle of the SM defined by $\sin\theta_W = g_1/\sqrt{g_1^2+g_2^2}$.
The gauge boson, $W^\pm$, is the charged weak gauge boson of the SM, and $\tilde{Z}$ and $\tilde{A}$ correspond to the $Z$ boson and 
photon in the SM limit, $(g',\epsilon) \to 0$.
In the following, we parameterize the VEVs as,
\begin{align}
v_1 = v \sin\beta,~v_2 = v \cos\beta,~v^2 = v_1^2 + v_2^2. \label{eq:beta}
\end{align}
With this parametrization, the charged weak gauge boson mass is given by
\begin{align}
m_W = \frac{g_2}{2} v.
\end{align}
The mass terms of the neutral gauge bosons can be casted in a $3 \times 3$ matrix as
\begin{align}
\mathcal{L}_{\mathrm{mass}} &= \frac{1}{2} \tilde{F}^T_\mu m_{\tilde{F}}^2 \tilde{F}^\mu,
\end{align}
where $\tilde{F}_\mu = (\tilde{A}_\mu, \tilde{Z}_\mu, \tilde{X}_\mu)^T$, and $m_{\tilde{F}}^2$ is given by
\begin{align}
m_{\tilde{F}}^2 &= 
\begin{pmatrix}
0 & 0 & 0 \\
0 & m_{\tilde{Z}}^2 & - g' v m_{\tilde{Z}} (q_1 \sin^2 \beta + q_2 \cos^2\beta) \\
0 & - g' v m_{\tilde{Z}} (q_1 \sin^2 \beta + q_2 \cos^2\beta) & g'^2 v_s^2 + g'^2 v^2 (q_1^2 \sin^2 \beta + q_2^2 \cos^2\beta )\\
\end{pmatrix}.
\end{align}
Here, $m_{\tilde{Z}}$ is the SM $Z$ boson mass defined by
\begin{align}
m_{\tilde{Z}} = \frac{1}{2} \sqrt{g_1^2 + g_2^2} v.
\end{align}
To obtain the masses of the neutral gauge bosons, we first diagonalize the gauge boson kinetic term by changing the basis of the fields 
$\tilde{F}$ to $\overline{F}=(\overline{A}, \overline{Z}, \overline{X})^T$ as  
\begin{align}
\tilde{F} &= U_K \overline{F},
\end{align}
where $U_K$ is an orthogonal matrix given by
\begin{align}
U_K &= 
	\begin{pmatrix}
	 1 & 0 & \epsilon r \cos\theta_W \\
	 0 & 1 & - \epsilon r \sin\theta_W \\
	 0 & 0 & r 
	\end{pmatrix},
\end{align}
with $r=(1-\epsilon^2)^{-1/2}$. Then, the mass matrix in $\overline{F}_\mu$ basis is given as
\begin{align}
m_{\overline{F}}^2  = U_K^T m_{\tilde{F}}^2 U_K 
= 
\begin{pmatrix}
0 & 0 & 0 \\
0 & m_{\tilde{Z}}^2 & -r m_{\tilde{Z}} \delta_1 \\
0 & -r m_{\tilde{Z}} \delta_1 & r^2 (g'^2 v_s^2 + \delta_1^2 + \delta_2^2)
\end{pmatrix},
\label{eq:mass-matrix2}
\end{align}
with
\begin{align}
\delta_1 &= \epsilon \sin\theta_W m_{\tilde{Z}} + g' v (q_1 \sin^2\beta  + q_2 \cos^2\beta),\\
\delta_2 &= |g' v (q_1- q_2)\sin\beta  \cos\beta |.
\end{align}
Next, the mass matrix Eq.\eqref{eq:mass-matrix2} can be diagonalized by an orthogonal matrix $V_F$
\begin{subequations}
\begin{align}
\overline{F} &= V_F F, \\
V_F &= 
\begin{pmatrix}
1 & 0 & 0 \\
0 & \cos\chi & -\sin\chi \\
0 & \sin\chi & \cos\chi
\end{pmatrix},
\end{align}
\end{subequations}
where $F = (A, Z, X)^T$ is the mass eigenstates. Their mass eigenvalues are given by 
\begin{subequations}
\begin{align}
m_A^2 &= 0,  \\
m_Z^2 &= 
m_{\tilde{Z}}^2 \cos^2\chi + r^2 (g'^2 v_s^2 + \delta_1^2 + \delta_2^2) \sin^2\chi 
 - 2 r m_{\tilde{Z}} \delta_1 \sin \chi \cos \chi,  \label{eq:z-mass} \\
m_X^2 &= 
r^2 ( g'^2 v_s^2 + \delta_1^2 + \delta_2^2) \cos^2\chi + m_{\tilde{Z}}^2 \sin^2\chi 
+ 2 r m_{\tilde{Z}} \delta_1 \sin \chi \cos \chi.  \label{eq:x-mass}
\end{align}
\label{eq:gauge-masses}
\end{subequations}
The mixing angle $\chi$ can be expressed as 
\begin{align}
\tan \chi = - \frac{r m_{\tilde{Z}} \delta_1}{m_{\tilde{Z}}^2 - m_X^2}. \label{eq:chi}
\end{align}
Here the mass of $X$ is an input of the model which should be $\simeq 17$ MeV by the Atomki experiment. 
In the situation of $m_{\tilde{Z}} \gg m_X$, the leading term of Eq.~\eqref{eq:chi} is given by
\begin{align}
\tan\chi \simeq -r \epsilon \sin\theta_W - r \frac{g' v}{m_{\tilde{Z}}} (q_1 \sin^2\beta + q_2 \cos^2\beta). \label{eq:tan-chi2}
\end{align}
In the parameter space of our interest, $g'$ and $\epsilon$ are roughly $\mathcal{O}(10^{-4}-10^{-3})$. 
Therefore, $\chi$ is much smaller than unity from Eq.~\eqref{eq:tan-chi2}.
Then, the difference between $m_Z$ and $m_{\tilde{Z}}$ is roughly given as, 
\begin{align}
m_{Z}^2 - m_{\tilde{Z}}^2 \simeq \delta_1^2 \simeq \mathrm{max}(\epsilon^2 m_{\tilde{Z}}^2, g'^2v^2) 
\sim (100~\mathrm{MeV})^2,
\end{align}
where Eq.~\eqref{eq:z-mass} is used.
This difference is smaller than the present error of the measured $Z$ boson mass, $91.1876 \pm 0.0021$ GeV \cite{Tanabashi:2018oca} and 
therefore we use $m_{\tilde{Z}} \simeq m_{Z} = 91.1876$ GeV as an input value in the following discussion.
Then, $v_s$ is expressed in terms of other parameters as
\begin{align}
v_s^2 = \frac{m_{X}^2 (m_Z^2 + r^2 (\delta_1^2 + \delta_2^2)) - r^2 m_Z^2 \delta_2^2 - m_X^4}{r^2 g'^2 (m_Z^2 - m_X^2)}.
\label{eq:vs-sq}
\end{align}
Right-hand-side of Eq.~\eqref{eq:vs-sq} should be positive for consistency.

In the end, the gauge eigenstates are expressed in terms of the mass eigenstates as 
\begin{align}
\begin{pmatrix}
\tilde{A}_\mu \\ 
\tilde{Z} _\mu\\ 
\tilde{X}_\mu
\end{pmatrix}
= U
\begin{pmatrix}
A_\mu \\ 
Z _\mu\\ 
X_\mu
\end{pmatrix} 
& =
\begin{pmatrix}
A_\mu + U_{12} Z_\mu +  U_{13} X_\mu \\
U_{22} Z_\mu + U_{23} X_\mu \\
U_{32} Z_\mu + U_{33} X_\mu
\end{pmatrix},
\label{eq:gauge-mixing}
\end{align}
where $U = U_K V_F$ and its elements are 
\begin{subequations}
\begin{align}
	U_{12} &= \epsilon r \cos\theta_W \sin\chi, ~~~~~~~~~~~~~~U_{13} = \epsilon r \cos\theta_W \sin\chi, \\
	U_{22} &= \cos \chi - \epsilon r \sin\theta_W \sin\chi, ~~~~U_{23} = -\sin\chi - \epsilon r \sin\theta_W \cos\chi, \\
	U_{32} &= r \sin\chi, ~~~~~~~~~~~~~~~~~~~~~~~~ U_{33} = r \cos\chi.
\end{align}
\label{eq:matrix-U}
\end{subequations}
From Eqs.~\eqref{eq:sm-gauge-bosons} and \eqref{eq:gauge-mixing}, the Lagrangian can be written in the mass basis of the gauge boson.

%%%%%%%%%%%%%%%%%%%%%%%%%%%%%%%%%%
\section{Couplings of the $X$ boson to fermions} \label{sec:coupling}
%%%%%%%%%%%%%%%%%%%%%%%%%%%%%%%%%%
In this section, we present the coupling constants of the $X$ boson to quarks and leptons.
The gauge interactions of the fermions to the $X$ boson are modified due to the mixing among the gauge bosons. 
Using Eqs.~\eqref{eq:sm-gauge-bosons} and \eqref{eq:gauge-mixing} with Eqs.~\eqref{eq:matrix-U}, 
the interaction Lagrangian of fermions, $f~(=u,d,\nu, N)$, can be written in the following form, 
\begin{align}
\mathcal{L}_{\mathrm{int}} 
= e \overline{f} \gamma^\mu ( \epsilon_f^V + \epsilon_f^A \gamma^5) f X_\mu, \label{eq:gauge-int}
\end{align}  
where $e$ is the proton electric charge. The vector coupling $\epsilon_f^V$ and axial-vector couplings $\epsilon_f^A$ 
are given by
\begin{subequations}
\begin{align}
\epsilon_u^V &= \frac{1}{4} \epsilon_R r \cos\chi + \frac{2}{3} \epsilon r \cos\theta_W \cos\chi 
			- \left( \frac{1}{4} - \frac{2}{3} \sin^2\theta_W \right) \epsilon_{\mathrm{NC}}, \\
\epsilon_u^A &= \frac{1}{4} \epsilon_R r \cos\chi + \frac{1}{4} \epsilon_{\mathrm{NC}}, \label{eq:epsUA}\\
\epsilon_d^V &= - \frac{1}{4} \epsilon_R r \cos\chi - \frac{1}{3} \epsilon r \cos\theta_W \cos\chi 
			+ \left( \frac{1}{4} - \frac{1}{3} \sin^2\theta_W \right) \epsilon_{\mathrm{NC}}, \\
\epsilon_d^A  &= -\epsilon_u^A =  \epsilon_e^A 
			= - \frac{1}{4} \epsilon_R r \cos\chi - \frac{1}{4} \epsilon_{\mathrm{NC}}, \label{eq:epsDA}\\
\epsilon_e^V &= - \frac{1}{4} \epsilon_R r \cos\chi - \epsilon r \cos\theta_W \cos\chi 
			+ \left( \frac{1}{4} - \sin^2\theta_W \right) \epsilon_{\mathrm{NC}}, \\
\epsilon_\nu^V &= - \epsilon_\nu^A = - \frac{1}{4} \epsilon_{\mathrm{NC}}, ~~
\epsilon_N^V = \epsilon_N^A =  \frac{1}{4} \epsilon_R r c_\chi,
\end{align}
\label{eq:fermion-gauge-coup}
\end{subequations}
with $\epsilon_R = g'/e$, where $\epsilon_{\mathrm{NC}}$ represents the neutral current contribution defined by
\begin{align}
\epsilon_{\mathrm{NC}} = \frac{\sin\chi + \epsilon r \sin\theta_W \cos\chi}{\sin\theta_W \cos\theta_W}. 
\label{eq:epsNC}
\end{align}
In Eqs.~\eqref{eq:fermion-gauge-coup}, we neglect the mixing between left and right handed neutrinos.\footnote{The mixing between 
the left and right-handed neutrinos is roughly given by $\sqrt{\frac{m_\nu}{M}}$ where $m_\nu$ and $M$ are the active neutrino mass  and Majorana mass, respectively. Taking $Y_N = \mathcal{O}(1)$, The Majorana mass is $\mathcal{O}(v_s)$ and larger than $10$ GeV for $g' < 10^{-3}$. Thus, the mixing is smaller than $10^{-4}$ for $m_\nu \sim 0.1$ eV. }
As we explained above, one of the most stringent constraints comes from neutrino-electron scattering of reactor neutrinos measured at TEXONO \cite{Deniz:2012fi}.
The left-handed neutrinos $\nu_L$ can interact with the $X$ boson through the weak neutral current. 
Thus, the coupling constant of $\nu_L$ is proportional to $\epsilon_{\mathrm{NC}}$ as 
\begin{align}
\epsilon_{\nu_L} = -\frac{1}{2} \epsilon_{\mathrm{NC}}. \label{eq:epsNuL}
\end{align}

To obtain approximate formulae of the coupling constants, we expand $\epsilon_{\mathrm{NC}}$ in the limit of 
$|\chi| \ll 1$ and $|Q| \ll 1$ as, 
\begin{align}
\epsilon_{\mathrm{NC}} 
& \simeq -\frac{m_{\tilde{Z}}^2 Q \epsilon_R \cos \theta_W + \epsilon m_X^2}{\cos \theta_W (m_{\tilde{Z}}^2 - m_X^2)}  \nonumber \\
& \simeq - Q \epsilon_R 
- \left( 
 Q\epsilon_R + \frac{\epsilon}{\cos\theta_W} 
\right) \frac{m_X^2}{m_{\tilde{Z}}^2} 
+ \mathcal{O}\left( 
Q \epsilon_R \frac{m_X^4}{m_{\tilde{Z}}^4},~
\frac{\epsilon}{\cos\theta_W} \frac{m_X^4}{m_{\tilde{Z}}^4}
\right), \label{eq:nc-leading}
\end{align}
where we define $Q$ for convenience as
\begin{align}
Q = (q_1 + q_2) - (q_1 - q_2) \cos 2\beta.
\end{align}
In the expansion, we kept the leading term of $\epsilon$ and $\epsilon_R$ and neglected higher order terms 
of these couplings  in each power of $m_X^2/m_{\tilde{Z}}^2$. Inserting Eq.~\eqref{eq:nc-leading} 
into Eqs.~\eqref{eq:fermion-gauge-coup}, the approximate expression of the coupling constants can be obtained, 
which is useful to understand signal requirement and constraints as we will explain later.

The first term of Eq.~\eqref{eq:nc-leading} vanishes when $\beta$ takes a specific value of
\begin{align}
\cos2\beta^\ast \equiv \frac{q_1 + q_2}{q_1-q_2}. \label{eq:beta-ast}
\end{align}
Then, the remaing term is much smaller than $\epsilon$ due to $m_{\tilde{Z}} \gg m_X$.
From Table \ref{tab:model}, $\cos2\beta^\ast$ is given in each Model by
\begin{subequations}
\begin{align}
\mathrm{Model~1} &:~~\cos2\beta^\ast  = 0, \\
\mathrm{Model~2} &:~~\cos2\beta^\ast  = -2, \\
\mathrm{Model~3} &:~~\cos2\beta^\ast  = - \frac{1}{2}, \\
\mathrm{Model~4} &:~~\cos2\beta^\ast  = - \frac{3}{2}.
\end{align}
\end{subequations}
Thus, $Q$ can be vanished in Model $1$ and $3$, while there are no solutions for $Q=0$ in Model $2$ and $4$.
%%%%%%%%%%%%%%%%%%%%%%%%%%%%%%%%%%%%%%%
\begin{figure}[t]
\vspace{-1cm}
\begin{tabular}{c}
 \includegraphics[width=0.6\textwidth]{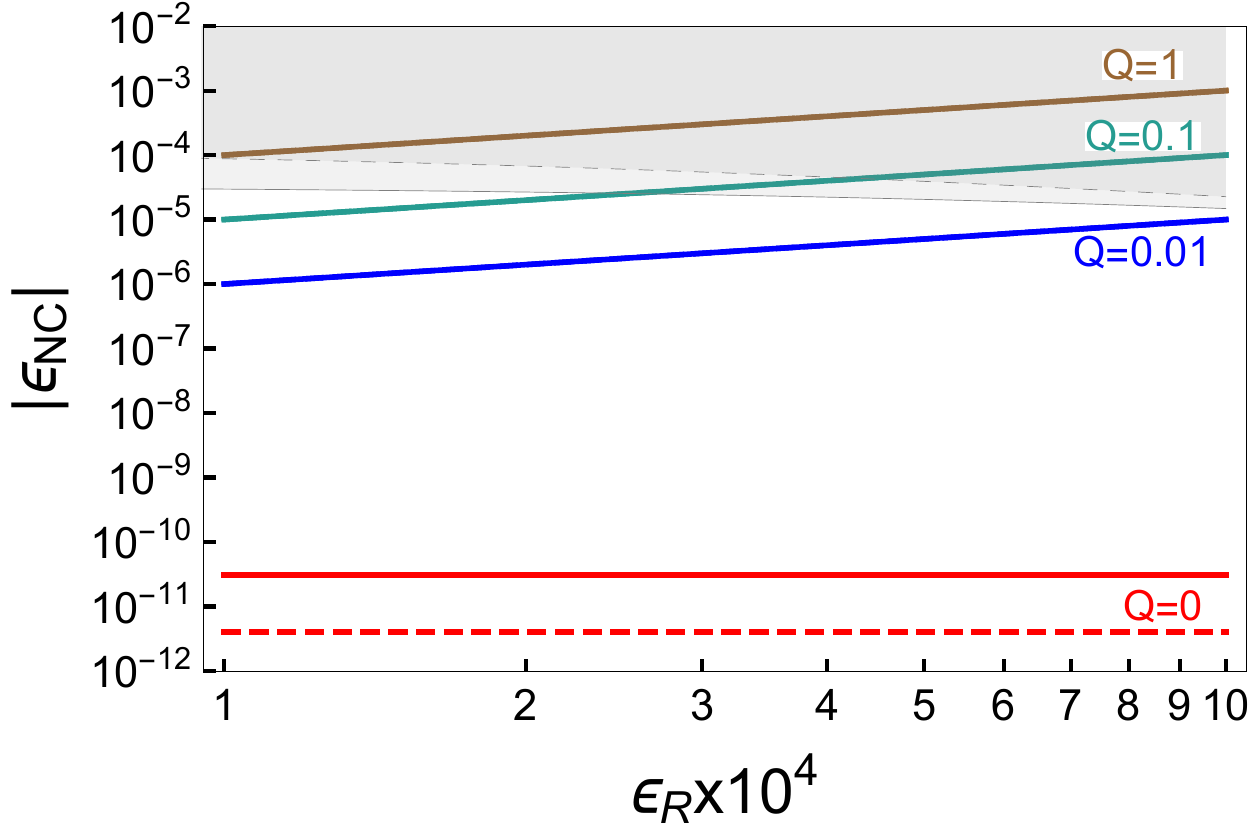} 
\end{tabular}
\caption{The coupling $\epsilon_{\mathrm{NC}}$ as a function of $\epsilon_R$. 
Red, blue, green and brown lines correspond to $Q=0,~0.01,~0.1$ and $1$, respectively. 
Solid and dashed ones correspond to $\epsilon = 5 \times 10^{-4}$ and $10^{-4}$. 
Gray filled region are exclusion region by the TEXONO results.}
\label{fig:epsilon-nc}
\end{figure}
%%%%%%%%%%%%%%%%%%%%%%%%%%%%%%%%%%%%%%%
Figure \ref{fig:epsilon-nc} is a plot of $|\epsilon_{\mathrm{NC}}|$ as a function of $\epsilon_R$. 
Red, blue, green and brown lines correspond to $Q=0,~0.01,~0.1$ and $1$, respectively. 
Solid and dashed ones correspond to $\epsilon = 10^{-4}$ and $5 \times 10^{-4}$. 
Gray filled regions with solid and dashed edges are exclusion region by neutrino-electron scattering for 
$\epsilon = 5 \times 10^{-4}$ and $10^{-4}$, 
which we will explain in subsection \ref{subsec:nu-e-scatt}.\footnote{It should be noted that 
the exclusion region in $\epsilon_R$-$\epsilon_{\mathrm{NC}}$ plane is almost independent of $Q$. 
In figure, we fixed $Q=0$. } 
Except for $Q = 0$, the dashed and solid curves are almost the same. 
One can see that the allowed region exists for $Q \leq 0.1$ while it does not for $Q \geq 1$. 
Thus, Model $1$ and $3$ can be evaded the constraint from $\nu$-$e$ scattering while Model 2 and 4, with $Q \geq 1$, 
are excluded for the choice of the parameters.

%%%%%%%%%%%%%%%%%%%%%%%%%%
\section{Signal and Experimental Constraints} \label{sec:signal-constraint}
%%%%%%%%%%%%%%%%%%%%%%%%%%
We summarize the signal requirement from the $^8$Be decay
\footnote{As we explained in the Introduction, the Atomki collaboration also reported that a peak like excess was found in $^4$He \cite{Krasznahorkay:2019lyl}, which can be consistently explained by 
a light particle for \Begr. However, nuclear matrix elements have significant uncertainty for $^4$He \cite{Feng:2020mbt} 
and we need further study to reduce the uncertainty. 
Thus we will not indicate $^4$He anomaly in our analysis.}
and the constraints from various experiments.

%%%%%%%%%%%%%%%%%%%%%%%%%%
\subsection{Signal Requirement}
%%%%%%%%%%%%%%%%%%%%%%%%%%

%%%%%%%%%%%%%%%%%%%%%%%%%%
\subsubsection{\Beex decay branching ratio}
%%%%%%%%%%%%%%%%%%%%%%%%%%
The Atomki collaboration has reported an anomalous internal pair creation for the M1 transition of the $18.15$ MeV excited 
state \Beex of \Begr \cite{Krasznahorkay:2015iga, Krasznahorkay:2017qfd,Krasznahorkay:2017gwn,
Krasznahorkay:2017bwh,Krasznahorkay:2018snd}.The collaboration measured angular correlations, and found a significant peak-like enhancement at larger 
angles. This result is mostly well-fitted under the assumption of the creation and subsequent decay 
of an intermediate particle $X$ with a mass of 
\begin{align}
m_X \simeq 17.0~\mathrm{MeV}.
\end{align}
%%%%%%%%%%%%%%%%%%%%%%%
\begin{table}[t]
  \begin{center}
    \begin{tabular}{|c|c|c|c|c|} \hline
    &~~~Previous Result~\cite{Krasznahorkay:2015iga}~~~&~~~~Exp1~~~~&~~~~Exp2~~~~&~~~Average~~~\\ \hline \hline
    $m_X$ (MeV)                       & $16.70(51)$ & $16.86(6)$ & $17.17(7)$ & $17.01(16)$ \\
    $B_X$  ($\times 10^{-6}$)    & $5.8$ & $6.8(10)$ & $4.7(21)$ & 6(1) \\
    ~~Significance~~     & $6.8\sigma$ & $7.37\sigma$ & $4.90\sigma$ & \\ \hline
    \end{tabular}
  \end{center}
\caption{The mass of $X$ particle and branching ratio of \Beex. }
\label{tab:signal}
\end{table}
%%%%%%%%%%%%%%%%%%%%%%%%
The signal branching ratio of \Beex into the assumed $X$ particle, followed by the decay of $X$ into $e^+e^-$,  
is defined by 
\begin{align}
B_X \equiv \frac{\Gamma(\mBeex \to \mBegr X)}{\Gamma(\mBeex \to \mBegr \gamma)}\mathrm{Br}(X \to e^+e^-),
\label{eq:br-X}
\end{align}
where $\Gamma(\mBeex \to \mBegr \gamma) \simeq (1.9 \pm 0.4)$ eV is the partial width of 
the $\gamma$ decay of 
\Beex and $\mathrm{Br}( X \to e^+ e^-)$ is the decay branching ratio of $X$ into an electron-positron pair.
From the Atomki experiment, the branching ratio \eqref{eq:br-X} and the $X$ boson mass have been 
constrained as given in Table \ref{tab:signal} (taken from \cite{ Krasznahorkay:2018snd}). 
These values have a relatively large uncertainties, which may originate from systematic uncertainty of 
unstable beam position in the experiment. Therefore, we employ rather conservative 
range for our numerical calculation
\begin{align}
4 \times 10^{-7} \lsim B_X \lsim 7 \times 10^{-6},
\label{eq:sig-req}
\end{align}
where $m_X$ is taken to $17.6$ and $16.7$ MeV for the lower and upper bounds, respectively.

To calculate the decay branching ratio of \Beex, we employ the results given in \cite{Kozaczuk:2016nma}. 
The partial decay width from the axial part of the gauge interaction is expressed as 
\begin{align}
\Gamma(\mBeex \to \mBegr X) 
= \frac{k}{18\pi}\,\left(2+\frac{E_k^2}{m_X^2}\right) \left |a_n \langle 0 || \sigma^n || 1 \rangle + a_p \langle 0 || \sigma^p || 1 \rangle \right| ^2,
\label{eq:partial-width}
\end{align}
where $k = \sqrt{\Delta E^2- m_X^2}$ and $E_k = \Delta E$ are the three momentum and energy of the $X$ boson, with 
$\Delta E = 18.15$ MeV being the difference of the energy level. The proton and neutron couplings, $a_p$ and $a_n$, are 
defined as
\begin{subequations}
\begin{align}
a_p &= \frac{a_0  + a_1}{2}, \label{eq:ap}\\
a_n &= \frac{a_0 - a_1}{2}, \label{eq:an}
\end{align}
\end{subequations}
where
\begin{subequations}
\begin{align}
a_0 &= (\Delta u^{(p)}+\Delta d^{(p)})(\epsilon_u^A + \epsilon_d^A) + 2  \Delta s^{(p)} \epsilon_d^A 
= 2  \Delta s^{(p)} \epsilon_d^A , \label{eq:a0} \\
a_1 &= (\Delta u^{(p)}-\Delta d^{(p)})(\epsilon_u^A - \epsilon_d^A)
= 2 (\Delta u^{(p)}-\Delta d^{(p)}) \epsilon_u^A, \label{eq:a1}
\end{align}
and $\epsilon_{u,d}^A$ is given in Eqs.~\eqref{eq:fermion-gauge-coup}.
\end{subequations}
The quark coefficients take values \cite{Bishara:2016hek}
\begin{subequations}
\begin{align}
\Delta u^{(p)} &= \Delta d^{(n)} ~=~ ~~~0.897(27),\\
\Delta d^{(p)} &= \Delta u^{(n)} ~=~ -0.367(27), \\
\Delta s^{(p)} &= \Delta s^{(n)} ~=~ -0.026(4),
\end{align}
\label{eq:quark-coeff}
\end{subequations}
and the nuclear matrix elements takes \cite{Kozaczuk:2016nma}
\begin{subequations}
\begin{align}
\langle 0^+ \| \sigma^p \| \mathcal{S} \rangle &=  -0.047 (29), \\
\langle 0^+ \| \sigma^n \| \mathcal{S} \rangle  &= -0.132 (33).
\end{align}
\label{eq:nme}
\end{subequations}
Inserting Eq.~\eqref{eq:partial-width} with these numbers into Eq.~\eqref{eq:br-X}, we obtain the branching ratio 
of \Beex decay.

Before closing this subsection, we show the parameter dependence of the signal branching ratio. 
Using $\epsilon_d^A = - \epsilon_u^A = \epsilon_e^A$, the partial decay width of \Beex is proportional 
to $(\epsilon_e^A)^2$ as
\begin{align}
\Gamma(\mBeex \to \mBegr X) \propto \frac{k}{18 \pi} \left( 2 + \frac{E_k^2}{m_X^2} \right) (\epsilon_e^A)^2. 
\label{eq:partial-width2}
\end{align}
The decay branching ratio of $X \to e^+ e^-$ is given by
\begin{align}
\mathrm{Br}(X \to e^+ e^-) = \frac{{\epsilon_e^V}^2 + {\epsilon_e^A}^2}{{\epsilon_e^V}^2 + {\epsilon_e^A}^2 + \frac{3}{2} \epsilon_{\nu_L}^2}, \label{eq:br-X-ee}
\end{align}
where $\epsilon_{e}^{V,A}$ and $\epsilon_{\nu_L}$ are given in Eqs.~\eqref{eq:fermion-gauge-coup} and \eqref{eq:epsNuL}. In Eq.~\eqref{eq:br-X-ee}, we have neglected the masses of electron and neutrino.
In the case of $|\chi| \ll 1$ and $|Q| \ll 1$, $\epsilon_e^A$ can be approximated as
\begin{align}
\epsilon_e^A \simeq -\frac{1}{4} (1 - Q) \epsilon_R 
+ \frac{1}{4} \left( Q \epsilon_R +\frac{\epsilon}{\cos\theta_W} \right) \frac{m_X^2}{m_{\tilde{Z}}^2}, 
\label{eq:approx-epsEA}
\end{align}
where Eq.~\eqref{eq:nc-leading} is used.
Similarly, the vector coupling of electron is approximated as
\begin{align}
\epsilon_e^V &\simeq - \frac{1}{4}( 1 + ( 1- 4\sin^2\theta_W)Q )\epsilon_R - \cos\theta_W \epsilon \nonumber \\
&\qquad -\left( \frac{1}{4} - \sin^2\theta_W \right) \left( Q \epsilon_R + \frac{\epsilon}{\cos\theta_W} \right)
\frac{m_X^2}{m_{\tilde{Z}}^2}, 
\label{eq:approx-epsEV}
\end{align}
Thus, neglecting $\mathcal{O}(m_X^2/m_{\tilde{Z}}^2)$ terms, the branching ratio, Eq.~\eqref{eq:br-X-ee}, 
is expressed by
\begin{align}
\mathrm{Br}(X \to e^+ e^-) \simeq 1 
- \frac{3}{2} \frac{4 Q^2 \epsilon_R^2}{(\epsilon_R + 4 \cos\theta_W \epsilon)^2 + \epsilon_R^2}. 
\label{eq:approx-br-X-ee}
\end{align}
%%%%%%%%%%%%%%%%%%%%%%%%%%%%%%%%%%%%%%%
\begin{figure}[t]
\vspace{-1cm}
\begin{tabular}{c}
 \includegraphics[width=0.6\textwidth]{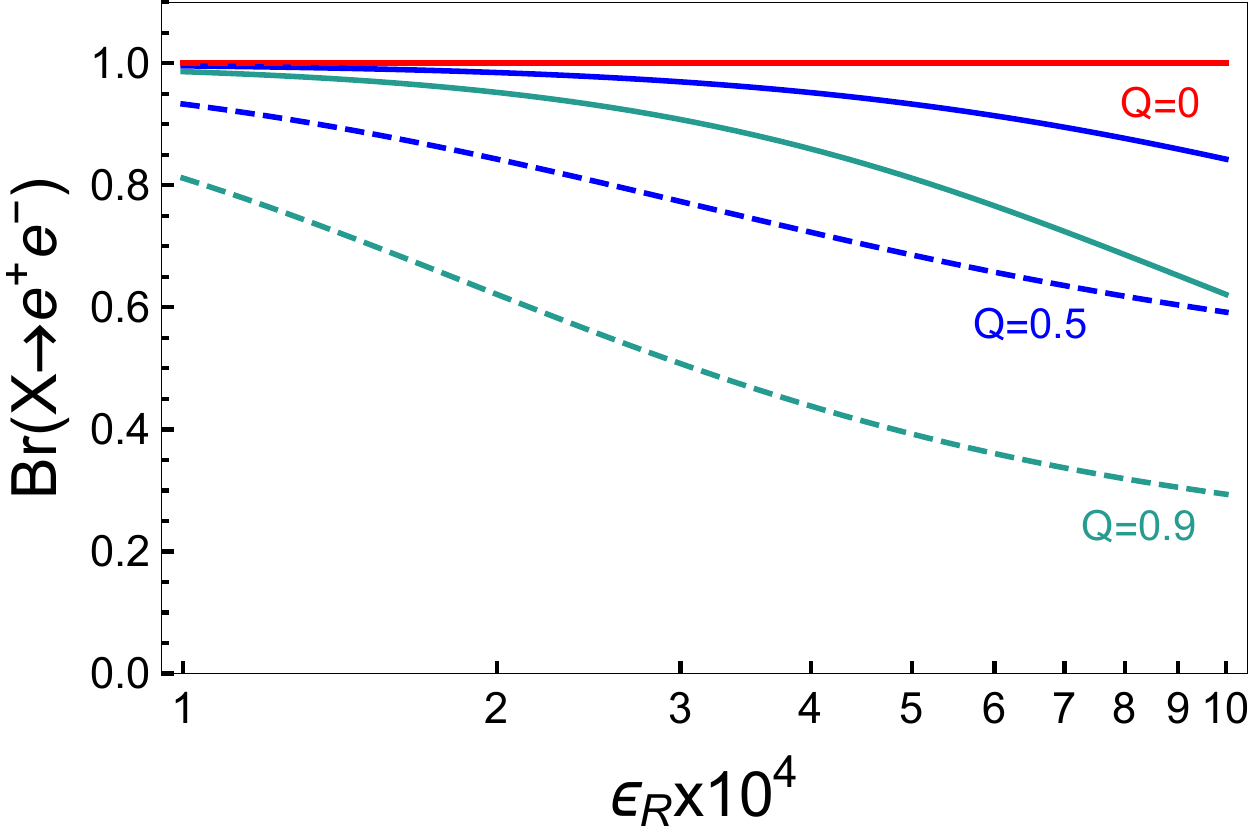} 
\end{tabular}
\caption{The branching ratio of $X \to e^+ e^-$ as a function of $\epsilon_R$. 
Red, blue and green curves correspond to $Q=0,~0.5$ and $0.9$, respectively. 
Solid and dashed ones correspond to $\epsilon = 5 \times 10^{-4}$ and $10^{-4}$. }
\label{fig:br-X-ee}
\end{figure}
%%%%%%%%%%%%%%%%%%%%%%%%%%%%%%%%%%%%%%%
Figure \ref{fig:br-X-ee} shows the branching ratio of $X \to e^+e^-$ as a function of $\epsilon_R$. 
We fixed as $\epsilon = 5 \times 10^{-4}$ (solid) and $10^{-4}$ (dashed) and as $Q=0$ (red), $0.5$ (blue) 
and $0.9$ (green), respectively. One can see that the branching ratio decrease as $Q$ increases. 
This is because the coupling to neutrinos, which is proportional to $\epsilon_{\mathrm{NC}}$, becomes large 
for non-vanishing $Q$ as shown in Fig.~\ref{fig:epsilon-nc}.

Using the approximate expressions, Eqs.~\eqref{eq:approx-epsEA} and \eqref{eq:approx-br-X-ee}, 
the signal branching ratio is scaled by the parameters as 
\begin{align}
B_X &\propto (\epsilon_e^A)^2 
\frac{{\epsilon_e^V}^2 + {\epsilon_e^A}^2}{{\epsilon_e^V}^2 + {\epsilon_e^A}^2 + \frac{3}{2} \epsilon_{\nu_L}^2} 
\nonumber \\
&\propto ( 1 - Q)^2 \epsilon_R^2 \left[  1 - \frac{3}{2} \frac{4 Q^2 \epsilon_R^2}{(\epsilon_R + 4 \cos\theta_W \epsilon)^2 + \epsilon_R^2} \right]. \label{eq:approx-br-X}
\end{align}
Thus, for $|Q| \ll 1$, the branching ratio is simply determined by $(\epsilon_R)^2$.

%%%%%%%%%%%%%%%%%%%%%%%%%%
\subsubsection{$X$ boson lifetime}
%%%%%%%%%%%%%%%%%%%%%%%%%%
To explain the Atomki anomaly, the new vector should decay inside the detector so that the electro-positron pair can 
be detected. As in \cite{Feng:2016ysn}, we require that $X$ boson propagates less than $1$ cm from its production point, 
which gives the condition as 
\begin{align}
\sqrt{ (\epsilon_e^V)^2 + (\epsilon_e^A)^2} \geq 1.3 \times 10^{-5} \sqrt{\mathrm{Br}(X \to e^+ e^-)}.
\end{align}

%%%%%%%%%%%%%%%%%%%%%%%%%%
\subsection{Constraints}
%%%%%%%%%%%%%%%%%%%%%%%%%%

%%%%%%%%%%%%%%%%%%%%%%%%%%
\subsubsection{Rare decay of neutral pion}
%%%%%%%%%%%%%%%%%%%%%%%%%%
The coupling constants of the light gauge boson to quarks can be constrained by meson decay experiments. 
The gauge boson can be produced in rare meson decays when those are kinematically allowed.
For the $X$ boson with $m_X \simeq 17$ MeV, the most stringent constraint among such meson decays comes from 
the rare decay of neutral pion into $X$ with a photon, i.e. $\pi^0 \to \gamma X$. 
Theoretically, only vectorial parts of the $X$ interaction to quarks can contribute to the decay.
The latest result of the NA48/2 experiment \cite{Raggi:2015noa} puts the following bound, 
\begin{align}
| 2 \epsilon_u^V + \epsilon_d^V | \leq \frac{0.3 \times 10^{-3}}{e \sqrt{\mathrm{Br}(X \to e^+ e^-)}}. \label{eq:const-p0}
\end{align}
The left-hand-side of the constraint is rewritten by
\begin{align}
2 \epsilon_u^V + \epsilon_d^V = - \epsilon_e^V,
\end{align}
and approximated as Eq.~\eqref{eq:approx-epsEV}. For $|Q| \ll 1$, Eq.~\eqref{eq:const-p0} is simplified as
\begin{align}
\left| \frac{1}{4} \epsilon_R + \cos\theta_W \epsilon \right| \leq 10^{-3}. \label{eq:approx-const-p0}
\end{align}

%%%%%%%%%%%%%%%%%%%%%%%%%%%%%%%%
\subsubsection{Neutrino-electron scattering} \label{subsec:nu-e-scatt}
%%%%%%%%%%%%%%%%%%%%%%%%%%%%%%%%
The interaction of the gauge boson to leptons, especially to neutrinos, is tightly constrained by 
neutrino-electron scattering \cite{Lindner:2018kjo}. 
The most stringent constraint for the $X$ boson 
is given by the TEXONO experiment \cite{Bilmis:2015lja}.
In \cite{Bilmis:2015lja}, the contributions from the $B-L$ gauge boson to $\overline{\nu_e}$-$e$ scattering 
have been studied. The authors analyzed the differential cross section with respect to the recoil energy of 
scattered electron and showed the interference term gives sizable contributions. 
Based on the analyses, the allowed region of the mass and gauge coupling of the $B-L$ gauge boson was shown. 
In this work, we derive the interference term of the differential cross section in the $U(1)_R$ model, and 
constrain the parameters by comparing the differential cross section in the SM
\footnote{Recently, in \cite{Hati:2020fzp}, this constraint is computed using data and $\chi^2$ 
fit is performed. As we have shown in Fig.~\ref{fig:epsilon-nc}, the coupling constant of neutrinos is 
much smaller than $10^{-5}$ for small $Q$. Therefore, our result is consistent with that of \cite{Hati:2020fzp}.
}.

The differential cross section in the SM and the interference term between the SM and $X$ boson contributions are given by
\begin{subequations}
\begin{align}
\left(\frac{d \sigma}{dT}\right)_{\mathrm{SM}} &= \frac{2 m_e G_F^2}{\pi E_\nu^2} 
\big( g_R^2 E_\nu^2 + g_L^2 (E_\nu - T)^2 -g_L g_R m_e T \big), \\
\left(\frac{d \sigma}{dT}\right)_{\mathrm{int}} &= \frac{g_\nu' m_e G_F}{\sqrt{2} \pi (m_X^2 + 2 m_e T)}
\big( 2 g_R g_L' E_\nu^2 + 2 g_L g_R' (E_\nu - T)^2 - (g_L g_R' + g_R g_L') m_e T\big) , 
\end{align}
\end{subequations}
where $T$ and $m_e$ are the recoil energy and the mass of electron, and $E_\nu$ is the energy of incident neutrino, respectively. 
The Fermi constant is denoted as $G_F$, and other coupling constants are given by 
\begin{subequations}
\begin{align}
g_L &= \frac{1}{2} + \sin^2\theta_W,~~~ g_R = \sin^2\theta_W, \\
g_L' &= e (\epsilon_e^V - \epsilon_e^A), ~~~ g_R' = e (\epsilon_e^V + \epsilon_e^A), \\
g_\nu ' &= 2 e \epsilon_\nu^V.
\end{align}
\end{subequations}
From \cite{Deniz:2012fi}, the event rate relative to that of the SM  is given by 
$1.08 \pm 0.21(\mathrm{stat}) \pm 0.16 (\mathrm{sys})$ in the TEXONO experiment. Thus we require 
\begin{align}
-0.64 < 
\frac{\left(\frac{d \sigma}{dT}\right)_{\mathrm{int}}}{\left(\frac{d \sigma}{dT}\right)_{\mathrm{SM}}} 
< 0.8, \label{eq:const-nu-e}
\end{align}
which corresponds to $3\sigma$ range. We use our numerical analysis $E_\nu = 3.0$ MeV and $T = 3.0$ MeV, 
respectively.

The ratio in Eq.~\eqref{eq:const-nu-e} can be approximated for the case of $|\chi| \ll 1$ and $|Q| \ll 1$ as,
\begin{align}
\frac{\left(\frac{d \sigma}{dT}\right)_{\mathrm{int}}}{\left(\frac{d \sigma}{dT}\right)_{\mathrm{SM}}}  \simeq 
-(5.0 \times 10^7 Q + 1.3) \epsilon \epsilon_R. \label{eq:approx-const-nu-e}
\end{align}
It can be understood from above equation that $Q$ is important for this constraint. 
Unless $Q$ is very close to zero, the constraint excludes 
$\epsilon \epsilon_R \gsim 10^{-8}/Q$.

%%%%%%%%%%%%%%%%%%%%%%%%%%%%%%%%%%
\subsubsection{Anomalous magnetic moment of charged lepton}
%%%%%%%%%%%%%%%%%%%%%%%%%%%%%%%%%%
The anomalous magnetic moment of the charged leptons have been measured accurately by 
experiments and also predicted precisely in the SM.
The new vector boson, that couples to the charged leptons, can shift the anomalous magnetic moments from the 
SM predictions via quantum loop corrections. One loop contribution of the vector boson is given by \cite{Fayet:2007ua}
\begin{align}
\delta a_l = \frac{e^2}{4 \pi^2}\left(  (\epsilon_l^V)^2 I_V(y_l) -  (\epsilon_l^A)^2  I_A(y_l) \right),
\end{align}
where $y_l=m_X^2/m_l^2$, and $I_V(y_l)$ and $I_A(y_l)$ are given by
\begin{subequations}
\begin{align}
I_V(y_l) &= \int_0^1 dx \frac{x^2 (1-x)}{x^2+(1-x) y_l}, \\
I_A(y_l) &= \frac{1}{y_l}\int_0^1 dx \frac{2 x^3 + (x-x^2) (4-x) y_l}{x^2 + (1-x) y_l}.
\end{align}
\label{eq:g-2-func}
\end{subequations}
It should be noticed that the axial coupling contribution to $\delta a_l$ is always negative while the vector contribution 
is positive. The integration of Eqs.~\eqref{eq:g-2-func} can be done numerically for electron and muon as 
\begin{subequations}
\begin{align}
I_V(y_e) &= 6.894 \times 10^{-7},~~~I_A(y_e) = 3.484 \times 10^{-6}, \label{eq:g-2-func-e}\\
I_V(y_\mu) &= 7.881 \times 10^{-4},~~~I_A(y_\mu) = 9.419 \times 10^{-2}, \label{eq:g-2-func-mu}
\end{align}
\end{subequations}
where we used $m_e = 0.5110$ MeV and $m_\mu = 105.7$ MeV, respectively. 

The muon anomalous magnetic moment has exhibited a long-standing discrepancy between experimental results 
\cite{Bennett:2006fi,Zlyla:2020xmw} and theoretical predictions 
\cite{Davier:2017zfy,Keshavarzi:2018mgv,Blum:2018mom,Davier:2019can}. 
From \cite{Zlyla:2020xmw}, the discrepancy is given by
\begin{align}
\Delta a_\mu = a_\mu^{\mathrm{exp}} - a_\mu^{\mathrm{SM}}  = (2.61 \pm 0.79) \times 10^{-9}, \label{eq:mu-g-2}
\end{align}
where $a_\mu^{\mathrm{exp}}$ and $a_\mu^{\mathrm{SM}}$ represent the anomalous magnetic moment by the 
measurements and SM predictions, respectively. 
From Eqs.~\eqref{eq:g-2-func-mu}, one finds that $\delta a_\mu$ can be positive when 
$|\epsilon_e^V| \gsim 10 |\epsilon_e^A|$. However, such parameter region is excluded by the constraint 
from $\pi^0 \to \gamma X$, because the lower bound on $\epsilon_e^A$ set by the signal requirement, 
\eqref{eq:approx-br-X}, results in too large $\epsilon_e^V$ in this situation. 
Then, the dominant part in $\delta a_\mu$ is the axial coupling term and that negative contribution 
to $\delta a_\mu$ further worsens the discrepancy. Thus a special care to implement the constraint 
of $\Delta a_\mu$ is required. Following the discussion in \cite{Kozaczuk:2016nma}, 
we impose a constraint that the contribution from the $X$ boson should be less than $2\sigma$ 
uncertainty of Eq.~\eqref{eq:mu-g-2},
\begin{align}
|\delta a_\mu| \lsim 1.58 \times 10^{-9}.\label{eq:const-mu-g-2}
\end{align}
This above constraint is scaled by the parameter as 
\begin{align}
|\delta a_\mu| = \frac{e^2}{64\pi^2}(1-Q)^2 I_A(y_\mu) \epsilon_R^2, \label{eq:approx-const-mu-g-2}
\end{align}
thus it is determined mostly by $\epsilon_R$.

The anomalous magnetic moment of electron, $a_e = (g-2)_e/2$, also has been measured accurately  
\cite{Hanneke:2008tm, Hanneke:2010au}, and predicted precisely within 
the SM \cite{Aoyama:2012wj,Aoyama:2014sxa, Laporta:2017okg,Aoyama:2017uqe}. 
Although recent results claimed that $a_e$ also exhibits $2.5\sigma$ discrepancy between the measurement 
and SM predictions, we impose a rather conservative constraint \cite{Giudice:2012ms} employed in \cite{Kozaczuk:2016nma},
\begin{align}
-26 \times 10^{-13} \lsim \delta a_e \lsim 8 \times 10^{-13}. \label{eq:delta-a-e}
\end{align}
This constraint is weaker than that from $\delta a_\mu$ due to the smaller value of $I_V$ and $I_A$.

%%%%%%%%%%%%%%%%%%%%%%%%%%%%%%%%
\subsubsection{Effective weak charge}
%%%%%%%%%%%%%%%%%%%%%%%%%%%%%%%%
The axial-vector coupling of electron can be restricted by atomic parity violation in Caesium (Cs) 
\cite{Bouchiat:2004sp,Davoudiasl:2012ag}. 
The constraint is given by the measurement of the effective weak charge $Q_W$ of 
the Cs atom \cite{Porsev:2009pr}. For the $X$ boson with $17$ MeV mass, one obtains the following constraint 
\begin{align}
\Delta Q_W = - \frac{2 \sqrt{2} e^2}{G_F} \epsilon_e^A \big[
(2Z+N) \epsilon_u^V + (Z+2N) \epsilon_d^V \
\big] 
\left( \frac{0.8}{(17.0~\mathrm{MeV})^2} \right) \leq 0.71,
\label{eq:effective-weak}
\end{align}
where $Z=58$ and $N=78$ are the number of proton and neutrino in Cs nucleus, respectively.

For $|Q| \ll 1$, $\Delta Q_W$ is given by
\begin{align}
\Delta Q_W \simeq - \frac{2 \sqrt{2} e^2}{G_F} \frac{\epsilon_R}{16} \big[
(Z+N) \epsilon_R + 4 Z \cos\theta_W \epsilon \
\big] 
\left( \frac{0.8}{(17.0~\mathrm{MeV})^2} \right) \leq 0.71. \label{eq:approx-const-weakcharge}
\end{align}

%%%%%%%%%%%%%%%%%%%%%%%%%%
\subsubsection{electron beam dump experiments}
%%%%%%%%%%%%%%%%%%%%%%%%%%
Another constraint is obtained by searches for gauge boson at electron beam dump experiments, such as 
SLAC E141 \cite{Riordan:1987aw,Bjorken:2009mm}, Orsay \cite{Davier:1989wz} and  NA64 \cite{Banerjee:2018vgk}, 
via bremsstrahlung from electron and nuclei scatterings.
The null results of these searches are interpreted as either (1) the gauge boson can not be produced due to 
very small coupling, or (2) the gauge boson decays rapidly in the dump.
For the $X$ boson to satisfy the Atomki signal requirement, the latter one restricts the electron couplings. 
From the latest result of NA64 \cite{Banerjee:2018vgk}, one obtain the constraint, 
\begin{align}
\sqrt{(\epsilon_e^V)^2 + (\epsilon_e^A)^2} \geq 4 \times 10^{-4}\sqrt{\mathrm{Br}(X \to e^+ e^-)}.
\end{align}
Using the approximate expression of $\epsilon_e^V$ and $\epsilon_e^A$, the constraint is given by
\begin{align}
(\epsilon_R + 4 \cos\theta_W \epsilon )^2 + \epsilon_R^2 \geq 2.6 \times 10^{-6}. \label{eq:approx-const-edump}
\end{align}

%%%%%%%%%%%%%%%%%%%%%%%%%%
\subsubsection{electron-positron collider experiments}
%%%%%%%%%%%%%%%%%%%%%%%%%%
The coupling to electron is also constrained by $e^+$-$e^-$ collider experiment such as KLOE-2 \cite{Anastasi:2015qla} 
and BaBar \cite{Lees:2014xha} experiments.
The most stringent limit on the $X$ boson has been set by KLOE-2, searching for $e^+ e^- \to \gamma X$ followed by $X \to e^+ e^-$,
\begin{align}
\sqrt{(\epsilon_e^V)^2 + (\epsilon_e^A)^2} \leq \frac{2 \times 10^{-3}}{\sqrt{\mathrm{Br}(X \to e^+ e^-)}}.
\end{align}
This constraint is weaker than that of electron beam dump experiment.

%%%%%%%%%%%%%%%%%%%%%%%%%%%%%%%%
\subsubsection{Parity violating M$\mathbf{\o}$ller scattering}
%%%%%%%%%%%%%%%%%%%%%%%%%%%%%%%%
Vector and axial-vector interactions of the $X$ boson to electrons induce an extra parity violation 
in M$\mathbf{\o}$ller scattering.
The cross section was measured at the SLAC E158 experiment \cite{Anthony:2005pm}.
The constraint for the $X$ boson with the mass of $17$ MeV is given as 
\begin{align}
|\epsilon^V_e \epsilon^A_e| \leq 1 \times 10^{-8}/e^2. \label{eq:moller}
\end{align}

%%%%%%%%%%%%%%%%%%%%%%%%%%
\subsubsection{vacuum expectation value of $S$}
%%%%%%%%%%%%%%%%%%%%%%%%%%
For consistency, Eq.~\eqref{eq:vs-sq} must be positive. Since $m_{\tilde{Z}}^2 \gg m_X^2$,  the requirement turns out to be
\begin{align}
m_{X}^2 (m_Z^2 + r^2 (\delta_1^2 + \delta_2^2)) - r^2 m_Z^2 \delta_2^2 - m_X^4 > 0. \label{eq:const-vev-s}
\end{align}
This constraint is also approximated as
\begin{align}
29 + 6.7 \epsilon^2 + 3.0 \times 10^8 q_2 \epsilon_R^2 > 0,
\end{align}
where we set $Q=0$ and $q_1 = 1/2$. Thus, the constraint excludes the parameter space when $q_2$ is negative. 
Assuming that $\epsilon$ is the same order of $\epsilon_R$, the exclusion region for $q_2 < 0$ is given by
\begin{align}
|\epsilon_R| > \frac{3.1 \times 10^{-4}}{\sqrt{|q_2|}}. \label{eq:approx-const-vev-s}
\end{align}

%%%%%%%%%%%%%%%%%%%
\section{Numerical Results} \label{sec:num-results}
%%%%%%%%%%%%%%%%%%%
In this section, we show our numerical results of the signal requirement and experimental constraints listed in the previous section. 
As we explained in section III, the coupling constant of left-handed neutrino in Model 2 and 4 is so large for any 
value of $\beta$ that the constraint from neutrino-electron scattering can not be evaded. We have analyzed the signal requirement and the constraints, and found no allowed region in these two models. 
Therefore, we show our numerical results on Model 1 and 3.

%%%%%%%%%%%%%%%%%%%
\subsection{Model 1}
%%%%%%%%%%%%%%%%%%%

%%%%%%%%%%%%%%%%%%%%%%%%%%%%%%%%%%%%%%%
\begin{figure}[t]
\vspace{-1cm}
\begin{tabular}{cc}
	\multicolumn{2}{c}{\includegraphics[width=0.5\textwidth]{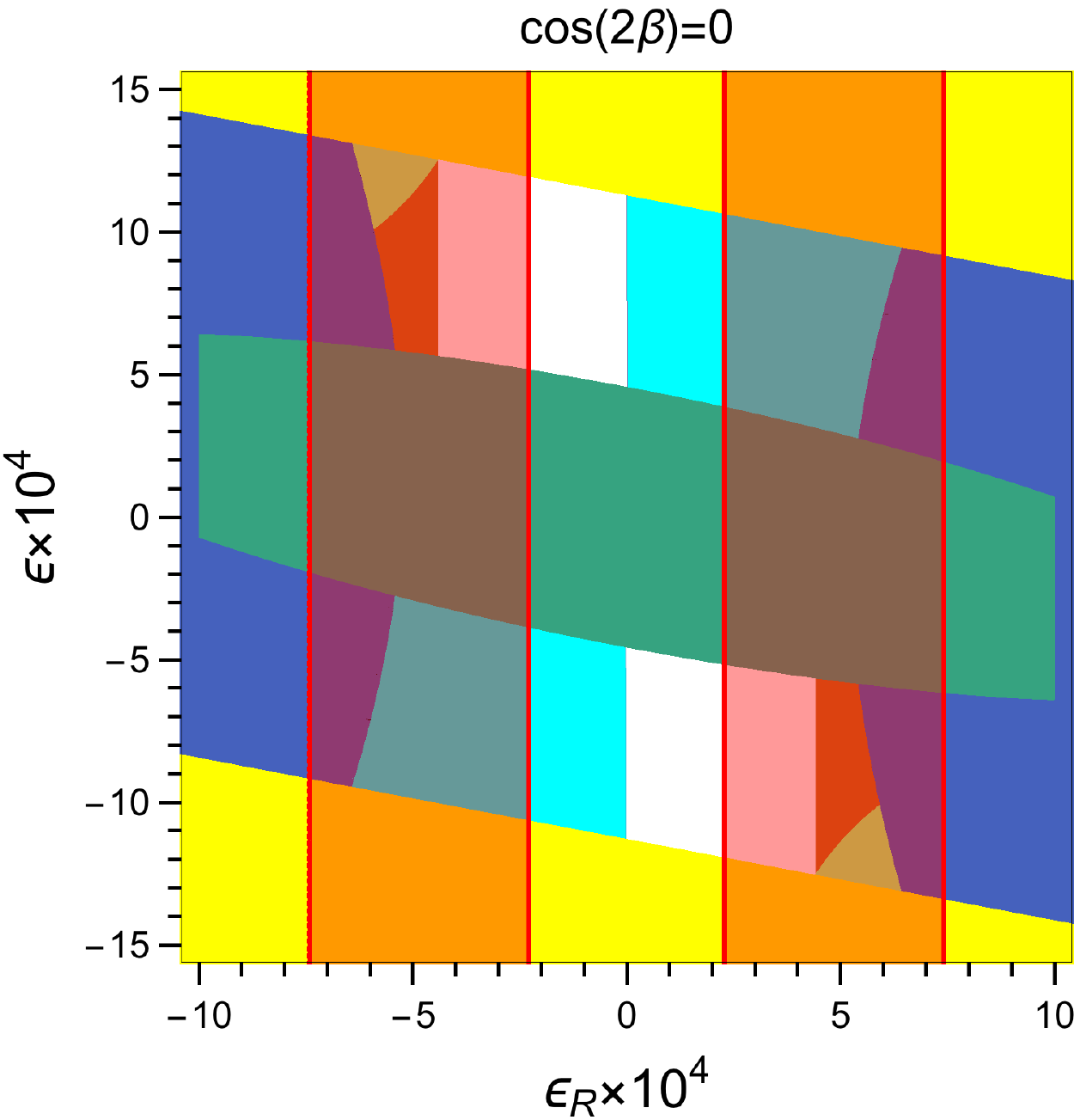}} \\
	\includegraphics[width=0.5\textwidth]{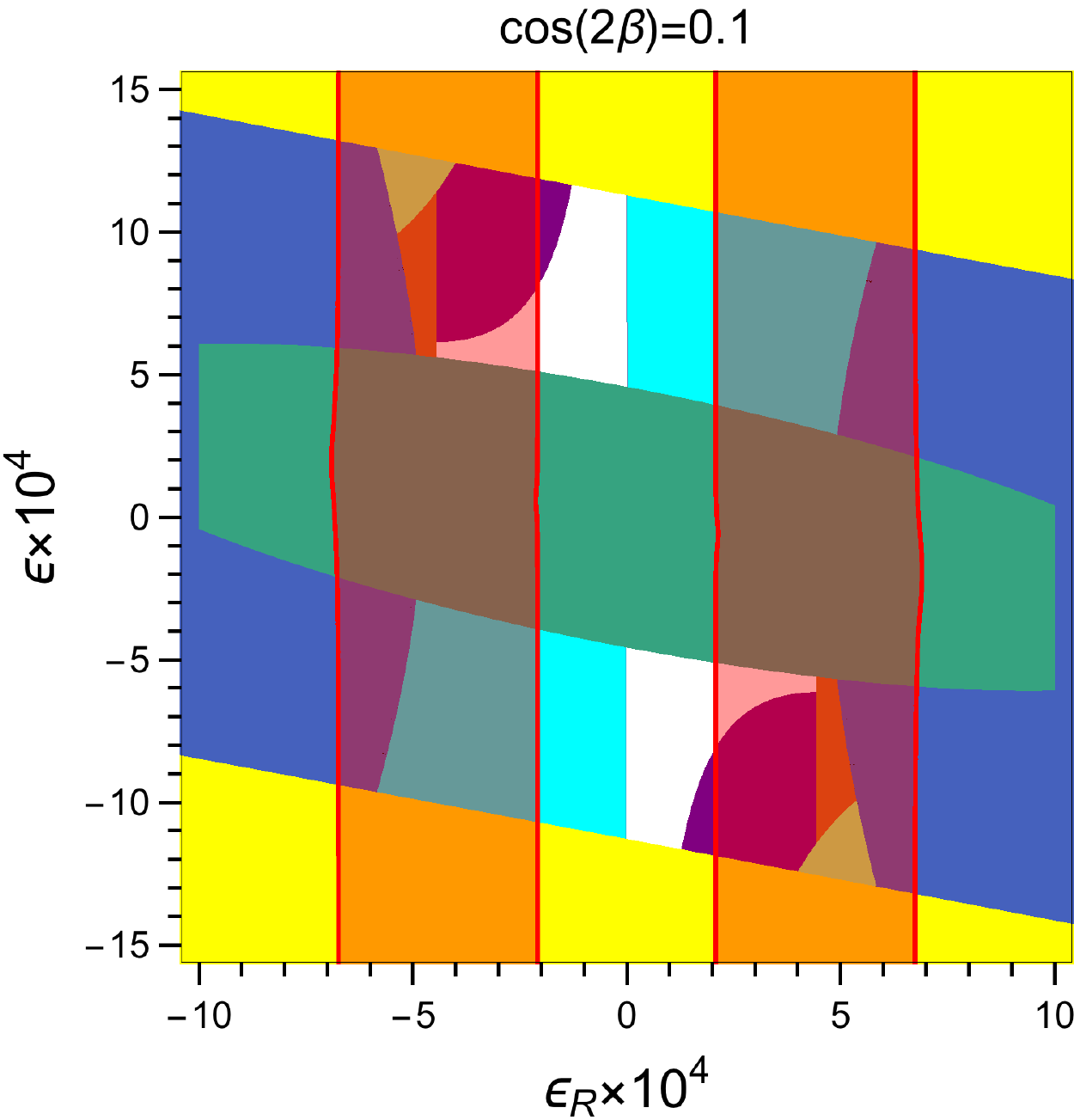} ~~&~~
	\includegraphics[width=0.5\textwidth]{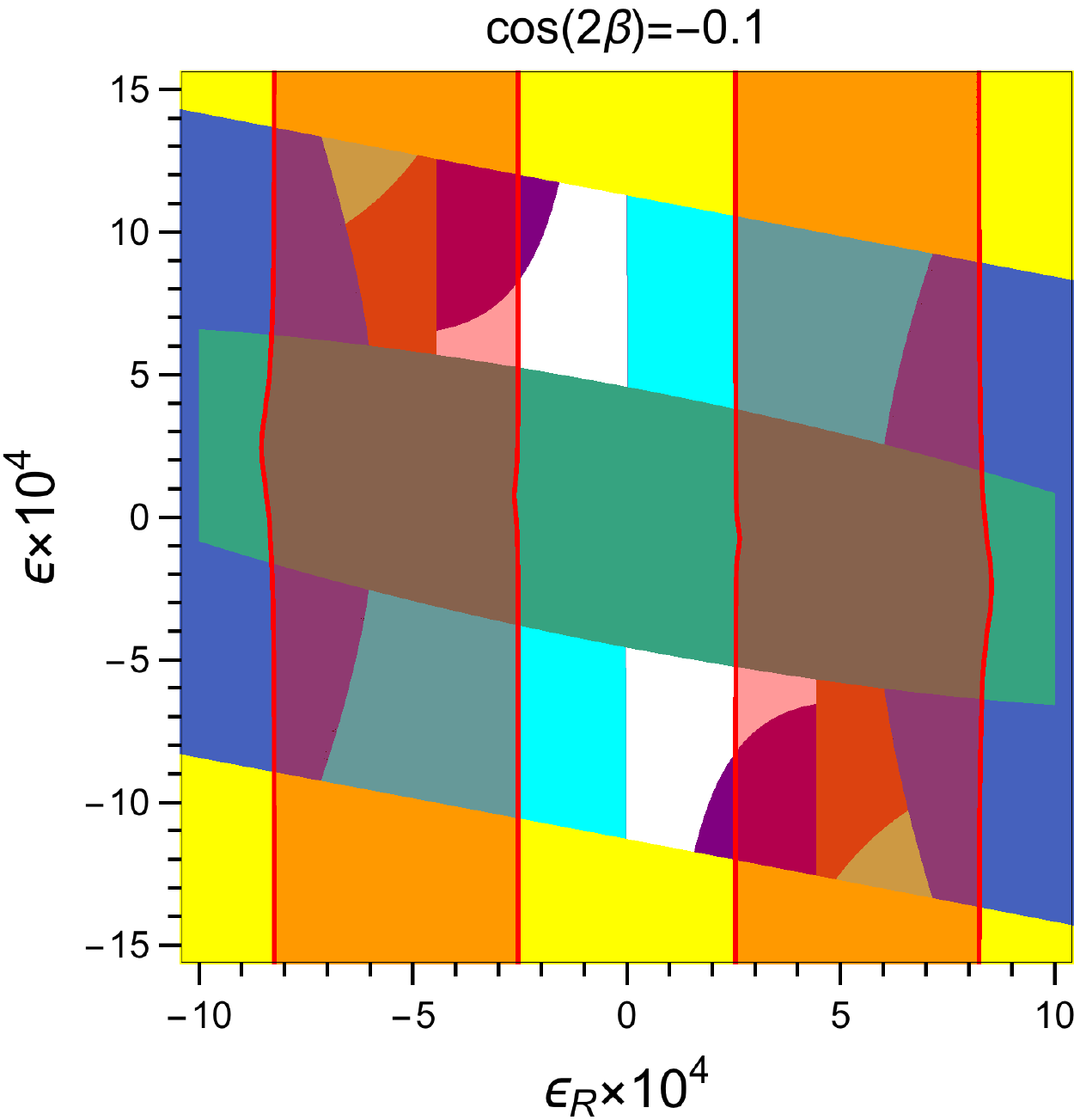} 
\end{tabular}
\caption{Allowed parameter region and signal in Model 1. 
The red band is the signal region 
for the Atomki results. White region is allowed by experiments, while other colored regions are excluded. See text for detail.
}
\label{fig:results-model-1}
\end{figure}
%%%%%%%%%%%%%%%%%%%%%%%%%%%%%%%%%%%%%%%
Figure \ref{fig:results-model-1} shows the signal region Eq.~\eqref{eq:sig-req} and exclusion regions 
for Model $1$ in $\epsilon_R$-$\epsilon$ plane. 
In this model, the factor $Q$, which determines the constraints from neutrino-electron scattering and signal, 
is given by, 
\begin{align}
Q = -\cos2\beta,
\end{align}
and hence $Q$ vanishes for $\cos2\beta = 0$.
We took $\cos 2\beta$ to $0$ and also $0.1$ for comparison as shown in the top of each panels. 
The mass of $X$ is fixed to be $17.01$ MeV for the constraints because the constraints are less sensitive to $m_X$, 
while it is taken to be $17.6$ and $16.7$ MeV for the signals, respectively.
Red transparent band represents the signal region with $B_X = 4 \times 10^{-7}$ (inside) 
and $7 \times 10^{-6}$(outside), respectively.  
Color filled regions are exclusion region by the experimental constraints from 
$\pi^0 \to \gamma X$ (yellow), 
$\nu$-$e$ scattering (purple), 
muon $g-2$ (dark blue), 
effective weak charge (light blue), 
electron beam dump experiment (green)
and also by the theoretical constraints from the positive VEV squared of the scalar field $S$ (brown). 

We first explain the general behavior of the signal requirement and constraints.
In each panel, one can see that the signal requirement is almost determined only by $\epsilon_R$, which 
is well approximated by Eq.~\eqref{eq:approx-br-X}.
For $Q=\pm 0.1$, the signal requirement shows slight dependence on $\epsilon$ in small $|\epsilon|$ region. 
In such region, the decay branching ratio of $X \to \nu \nu$ is not negligible, and therefore larger $\epsilon_R$ is 
needed to satisfy the signal by enhancing the decay of $X \to e^+ e^-$. 

About the constraints, one can also see that the $\pi^0 \to \gamma X$ constraint excludes the region in large $|\epsilon|$ 
while the constraints from $(g-2)_\mu$ and VEV of $S$ exclude the region in large $|\epsilon_R|$. 
The central region is excluded by the constraint from electron beam dump 
experiment. The constraint from effective weak charge excludes the region of $\epsilon \epsilon_R >0$. 
The qualitative behavior of these exclusion regions can be understood by the approximated expressions 
of the constraints, Eqs.~\eqref{eq:approx-const-p0},~\eqref{eq:approx-const-mu-g-2} 
and \eqref{eq:approx-const-edump},~\eqref{eq:approx-const-vev-s}.
The last one is the constraints from neutrino-electron scattering, which is well approximated by 
Eq.~\eqref{eq:approx-const-nu-e} with $\cos2\beta$ chosen here. 
It is seen that the constraint excludes large region of the parameter space in $\epsilon \epsilon_R < 0$ 
for $|\cos2\beta| = 0.1$, while it disappears for $\cos2\beta =0$.
This is because $\epsilon_{\mathrm{NC}}$ is not suppressed in the former cases.

For $\cos 2\beta = 0$ ($Q = 0$), we found wider parameter region 
consistent with the Atomki signal and all of the constraints. The coupling constants for this region is 
$2.2 \lsim |\epsilon_R| \times 10^{4} \lsim 4.4$, $5.1 \lsim |\epsilon| \times 10^{4} \lsim 12.4$.
For $|\cos 2\beta| = 0.1$, the consistent region in the parameter space is found in
$2.1 \lsim |\epsilon_R| \times 10^{4} \lsim 4.4$ and $5.1 \lsim |\epsilon| \times 10^{4} \lsim 8.0$, respectively.
For $|\cos 2\beta| \gsim 0.1$, there are no consistent region found in our analysis.

%%%%%%%%%%%%%%%%%%%
\subsection{Model 3}
%%%%%%%%%%%%%%%%%%%

%%%%%%%%%%%%%%%%%%%%%%%%%%%%%%%%%%%%%%%
\begin{figure}[t]
\vspace{-1cm}
\begin{tabular}{ccc}
& \includegraphics[width=0.5\textwidth]{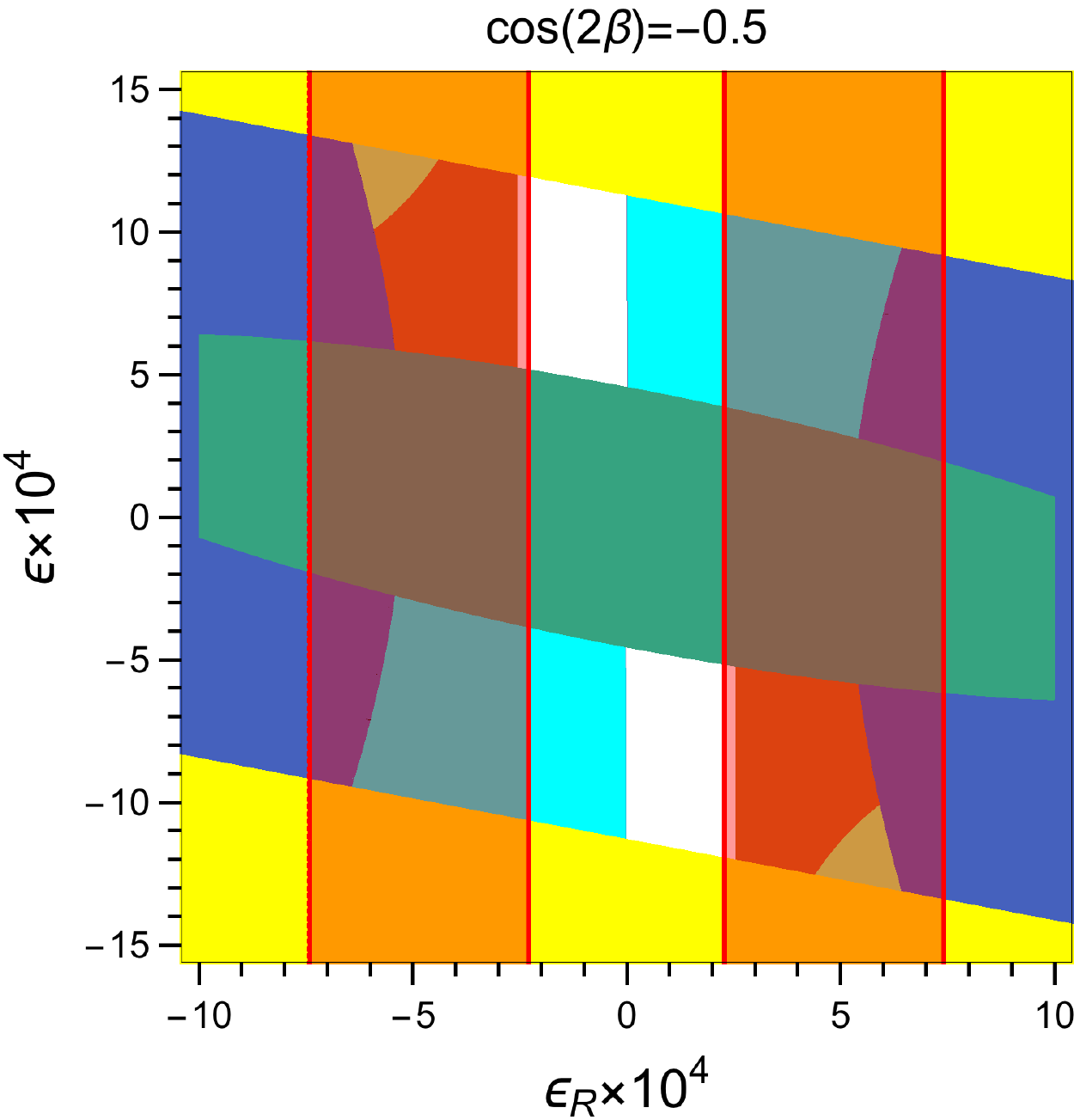} &
\end{tabular}
\caption{Allowed parameter region and signal in Model 3.}
\label{fig:results-model-3}
\end{figure}
%%%%%%%%%%%%%%%%%%%%%%%%%%%%%%%%%%%%%%%
Figure \ref{fig:results-model-3} is the same plot for model $3$ as Figure \ref{fig:results-model-1}. 
In this model, $Q$ is given by
\begin{align}
Q =  -1 - 2 \cos 2\beta,
 \end{align}
and hence we took $\cos2\beta = -0.5$.
The general behavior of the signal requirement and constraints is almost the same as in Model $1$. 
We found a narrow consistent region in the parameter space for $\cos2\beta= -0.5$. The coupling constant for 
this region is $2.3 \leq |\epsilon_R| \times 10^4 \leq 2.5$ and $5.1 \leq |\epsilon| \times 10^4 \leq 11.9$.
In this model, the constraint from the VEV of $S$ excludes most of the signal region, even if the neutrino-electron 
scattering constraint is avoided by taking $Q=0$. From Eq.~\eqref{eq:approx-const-vev-s}, the exclusion region 
from this constraint is obtained as 
\begin{align}
|\epsilon_R| > 2.5 \times 10^{-4},
\end{align} 
while that in Model $1$ is 
\begin{align}
|\epsilon_R| > 4.4 \times 10^{-4},
\end{align}
which is in good agreement with our numerical results.

%%%%%%%%%%%%%%%%%%%%%%%
\section{Conclusion} \label{sec:conclusion}
%%%%%%%%%%%%%%%%%%%%%%%
We have discussed the Atomki anomaly in the gauged $U(1)_R$ symmetric model. 
As a minimal model to solve the anomaly, three right-handed neutrinos are introduced for the cancellation of gauge 
anomalies. Two $SU(2)$ doublet and one SM singlet Higgs scalar particles are also introduced to evade 
the stringent constraints from neutrino-electron scattering and relativistic degree at the early Universe. 
Then, the new gauge boson is identified with the $X(17)$ boson. The Atomki signal requirement and 
other experimental constraints have been studied analytically and numerically in this model. 

We first classified models depending on the $U(1)_R$ charges of two doublet Higgs fields, by requiring all of  
CP-odd as well as CP-even scalars to be massive. We found that the possible choices of the gauge charges 
are limited to four cases $q_2 = -1/2,~\pm 3/2,~+5/2$. 
Two of them leads to large neutrino coupling to electron, and hence such cases are excluded by the constraint 
from neutrino-electron scattering. 
Then, for other models with $q_2 = -1/2$ and $-3/2$, called Model $1$ and $3$ respectively,  
we found that consistent regions with the signal and constraints exist. 
In such regions, the constraint from neutrino-electron scattering is suppressed due to the cancellation of the gauge charges 
between two Higgs doublets. 
In Model $1$, the consistent region can be found for $|\cos 2\beta| < 0.1$ and in Model $3$, it is found for $\cos 2\beta= -0.5$. 
Other values of $\cos\beta$ and also other models have been excluded by experimental and theoretical constraints.

{\it Comment:} While we were finishing this work, ref. \cite{Feng:2020mbt} appeared, in which the axial-vector hypothesis was examined. 
The authors concluded that $^8$Be and $^4$He anomalies could be explained by significant uncertainty in nuclear matrix elements.

\begin{acknowledgements}
This work is supported, in part, by JSPS KAKENHI Grant Nos.~19K03860 and 19K03865, and MEXT KAKENHI Grant No.~19H05091 (O.~S), 
and JSPS KAKENHI Grant Nos.~18K03651,~18H01210 and 
MEXT KAKENHI Grant No.~18H05543 (T.~S).
\end{acknowledgements}

%%%%%%%%%%%%%%%%%%
%%% references %%%
%%%%%%%%%%%%%%%%%%
\bibliographystyle{apsrev}
\bibliography{biblio}

\end{document}